\documentclass[%
 reprint,
 amsmath,amssymb,
 aps,
]{revtex4-1}

\usepackage{subfigure}
\usepackage{graphicx}
\usepackage{dcolumn}
\usepackage{bm}

\begin{document}

\preprint{APS/123-QED}

\title{The Structure And Charge Density Waves In $NbSe_2$  By First-Principles Calculations}

\author{Wending Zhao}
 \altaffiliation{Physics Department, Nanjing University.}
 \email{zhaohhwd@163.com}
\affiliation{%
Physics Department, Nanjing University, No.22 Hankou road, Nanjing, China
}%

\date{\today}

\begin{abstract}
In this paper, we investigated structures and charge density waves, including its origin of CDW, by using first-principles calculations. Firstly, we performed structure searches of $NbSe_2$ under different pressures to explore the possible structures of the CDW phase in $NbSe_2$. The stability of the resulted structure is proved by extensive studies of energy calculations and phonon spectra. Based on the detailed analysis on electronic properties (both in real space and K space) and band structures during CDW transitions, we believe the structure distortion and the displacement of atoms are the origin of CDW, which also cause the change of the Fermi surfaces and electronic density of states (DOS). In the end, the results of band structure and DOS were used to explain the reason why CDW and superconducting orders compete with each other at low temperature.

\begin{description}

\item[KEY WORDS]
Charge density waves, Perierls transition, first-principles calculation, Structure distortion, The layered transition metal dichalcogenides (TMDs)£¬ 2H-NbSe2£¬ Superconductor

\end{description}
\end{abstract}

\pacs{Valid PACS appear here}
\maketitle


\section{Introduction}
The layered transition metal dichalcogenides (TMDs), such as $2H-TaS_2$, $2H-NbSe_2$ and $2H-TiSe_2$, have attracted a great interesting, due to their electronic properties and phase transition\cite{Withers1986REVIEW}. Among a lot of fancy physical phenomenon in TMDs, charge density waves(CDW) transition becomes a top priority, especially. Among too many TMDCs with CDW transition, $2H-NbSe_2$ has gained greater attention and is investigated more as the first discovered and the most typical one in 2H-layer TMDs. In addition, CDW phase and superconductor(SC) phase can coexist and compete with each other in $2H-NbSe_2$ under low temperature.
Charge density wave is a collective state which often appears in 1D or 2D material, and it leads to the sinusoidal oscillation of charge density. Initially, CDW came from a concept, Perierls transition\cite{Berggren2010Electrons}£¬ which is periodic structure transition in one-dimensional (1D) system following with energy decreasing and periodic charge density waves\cite{Bardeen1990Superconductivity}. As Perierls transition saying, the interaction between electrons and phonon caused by structure distortion leads to charge density waves in one-dimensional system. In quasi 1D metal, electrons and phonons form intense nesting at Fermi level causing charge density waves at   for phonon¡¯s vector.
With the development of 2D layer material, more and more fancy phenomenon including CDW was discovered in 2H material. Besides, these behaviors and their origins are unclear and complicate in 2D system. One of them, $2H-NbSe_2$, is a successive Se-Nb-Se layer structure, and the interaction between layers is Van der Waals force. Charge density wave orders and Superconductor transition are the central issues attracting much attention in $2H-NbSe_2$. At $T_cdw=33.5K$, $2H-NbSe_2$  undergoes a second order phase transition from the normal phase to CDW phase, which have been found in $3¡Á3$ supercell by the result from STM and ARPES experiments, but there is not anyone experiment showing the periodicity on z-axis yet\cite{J2001Charge,Langer2014Giant}. Although $2H-NbSe_2$  is one of the early discovered CDW materials studied by experiment and theory extensively, the mechanism for CDW transition in the $2H-TMDC¡¯s$ is still a controversial topic\cite{M2007Fermi,Gr1994Density}. There are three different origins of CDW instability: fermi nesting\cite{Straub1999Charge}, saddle points\cite{Rice1975New}, and electron-phonon interactio\cite{Valla2000Charge,Kiss2007Charge,Weber2011Extended,C2015Quasiparticle}. The first mechanism is similar to the Peierls instability, however, only fractions of a fermi surface, whose electron¡¯s  density is larger than others, satisfy the required nesting condition for 2H-NbSe2. In the second one, the CDW vector is decided by k-space saddle points in the electronic structure close to the fermi level, Some other experiments and theories said the strong electron-phonon coupling is the major driving mechanism using density functional theory. When temperature is below 7.2K, CDW phase coexists with superconductivity in $2H-NbSe_2$, which has been studied in experiments\cite{Soto2006Electric} and theories\cite{Leroux2015Strong}. So the origin of CDW in $2H-NbSe_2 $ is still on debate, and it¡¯s hard to use first-principles computation to explain, because of defining structure of CDW order.
In this paper, we use structure searching to find structure distortion of CDW in $ 2H-NbSe2$, and explain the microscopic origin of CDW transition in $2H-NbSe_2$ by first-principles calculation. Then we will show the structure distortion and electron-phonon correlations are the origin of CDW by the electrical properties calculations of 2H-NbSe2. In the end, we will show the coexistence of CDW and SC phase by k-space electron density and band structure.

\section{Crystal Structure}
Before undergoing CDW transition, 2H-NbSe2 crystallizes in the space group $P6_3mmc$, and Nb atoms occupy $2b:(0,0,1/4), and (0,0,3/4) $and Se atoms occupy $4f: (1/3,2/3,z), (2/3,1/3,z+1/2), (2/3,1/3,-z), and (1/3,2/3,-z+1/2)] $sites, and the structure parameter is $a_0=3.474,b_0=3.474,c_0=13.7167$,$ ¦Á=90^\circ,¦Â=90^\circ,¦Ã=120^\circ$,  and $ z= 0.118$. In progress of searching CDW state, we make the periodicity  tripled in all directions($a=3a_0,b=3b_0,c=2c_0$) at first. Then we set some kinds of atoms displacements artificially, and put several more stable structures into seeds used for doing structure searching in USPEX. In the next, we take out the best structure with lowest energy, as CDW state. In the end, we compare these two states and get the atoms¡¯ displacement. We find that the structure parameters and z change on base of $3*3*2 $supercell especially along z-axis and the atom displacement in real-space is periodic function along X or Y axis. And atoms are displaced from their origin places by a small displacement. After distortions, Se atoms located at $(1/3+\delta_s2,2/3, z) $, $(1/3-\delta_s1,2/3+\delta_s3,-z+1/2) $,  $(2/3,1/3+\delta_s2,-z) $ and $(2/3+\delta_s1,1/3+\delta_s3,z+1/2)$ sites, where $\delta_s1=0.004a, \delta_s2=0.003a, $ and $\delta_s3=0.002b$. Nb atoms located at $(0,0+\delta_N1,1/4), (0,0+\delta_N,3/4) $sites, where $\delta_N=0.02b$. Meanwhile, the structure parameter turns to be $a=10.4059, b=10.4046, c=26.9125, ¦Á=90.281^\circ, ¦Â=89.5977^\circ, ¦Ã=119.9901^\circ $after CDW transition.

\section{Technical Details }
There are two approximations to the exchange correlation funciotn use, the local density approximation (LDA)\cite{Burke1998Derivation}.and the generalized gradient approximation (GGA)\cite{perdew1996generalized}, Spin-orbit coupling and Van der Waals force are included in calculation. The charge properties calculations are performed by using VASP and MATLAB. The band structure and phonon spectrum are calculated by VASP, and their Brillouin zone mesh is almost $9\times9\times2$.
\section{Content}

\subsection{The structure of $NbSe_2$}

For ensuring the most stable origin phase of $NbSe_2$, we did static self-consistent calculations toNbSe2 with different space symmetry group found by structure searching under different environment pressure, such as $P3m1$, $P6_3mmc$, $r3mh$ .

 \begin{figure}
  \centering
  \includegraphics[width=3.5in, height=2in]{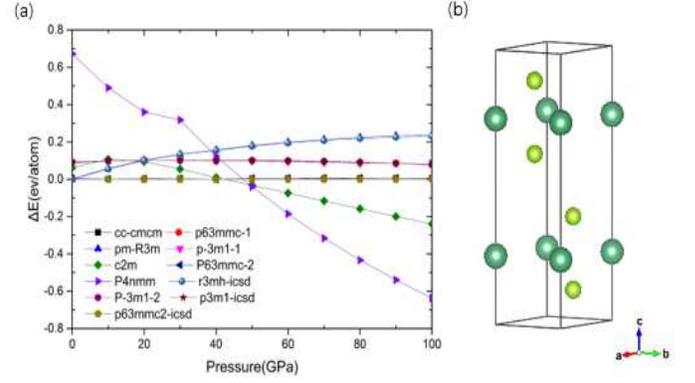}\\
  \caption{ (a) show the energy of $NbSe_2$ with different space group. The value on Y axis shows the difference between $P6_3mmc-icsd$, and x axis is pressure(GPa). Figure(b) is the schematic of $NbSe_2¡¯s$ structure with space group $P6_3mmc$, where the dark green points are Nb atoms, and light green points are Se atom. }
  \label{fig1}
\end{figure}
According to Fig.~\ref{fig1}, the structure with space group $P6_3mmc$has the lowest energy when the pressure is under $5GPa$. There are three different line which corresponding to which means it¡¯s the most stable structure in thermal. During the SCF calculation, we have taken Van der Waals forces into consideration which play an important role in this 2D material, $2H-NbSe2$, and the results are accurate enough.
This schematic (Fig.(1b)) shows the layer structure of $NbSe_2$, in which one layer of Nb atoms is sandwiched by two Se atoms layers. Its crystal structure parameters have been given before.
As we all known, $NbSe_2$ undergoes CDW transition when $T_CDW=33.5K$, going with structure distortion and the change of crystal¡¯s period . Thus, we make a structure searching where we use $P6_3/mmc$(from icsd database) and its supercells as seeds of searching to ensure the specific structure of $NbSe_2$ at low temperature. Finally, we got several new structures by USPEX, named as $P6_3/mmc-1$  and $P6_3/mmc-2$ respectively. More important, this two structures are $3\times3$ supercell of $NbSe_2$. And a more accurate SCF calculation was taken for this three structures when pressure is between $0$ and $1 GPa$,
 \begin{figure}
  \centering
  \includegraphics[width=3.5in, height=2in]{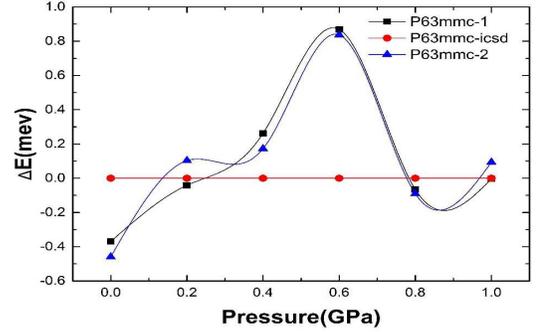}\\
  \caption{ Three different $NbSe_2's$  structures with $P6_3/mmc$ space group, X axis is pressure(GPa), and Y axis is the different energy value $\Delta E(meV)$  between $ P6_3/mmc-icsd¡¯s $}
  \label{fig2}
\end{figure}

According to Fig.~\ref{fig2}, this two new structure found in structure searching by USPEX have lower energy that the origin structure gotten from ICSD-database($P6_3/mmc-icsd$) at $0GPa$. And their energy will increase to the same value with the origin structure around $0.2 GPa$. Their energy will be more than the origin structure at $0.2-0.8 GPa$, and tend to be the same value with pressure increasing. Because the energy¡¯s changes are caused by structure distortion and atom¡¯s displacements whose difficulty are decided by pressure. With pressure increasing, structure distortion will be more difficult to happen. Although the different value of energy is only $10^(-4)(ev/atom)$, it was proven to be right that structure distortion and atoms¡¯ displacements can cause energy decreasing at low pressure. This point enlightens us to use structure distortion and atoms¡¯ displacements to search CDW order of $NbSe_2$. In fact, this idea is also identical to Perierls transition which points out the relationship between structure distortion and CDW transition.
So we use these two $3\times3$ supercell structures to investigate how structure is affected by CDW transition. About XY plane, there are a lot of papers and researches which have proven it is $3\times3$ period, which is identical to our results. As for z axis, there is no paper giving clear conclusion.
\subsection{The origin of CDW in $NbSe_2$}
According to the comparison between the supercell structure searched by USPEX and the origin structure, we get a most stable crystal structure, called CDW order. Then, we define this kind of structure distortion and atom displacement as the standard change mode which has been given in chapter 2. In the next, we make different multiples (such as $\times0.25, \times0.5, \times0.75, \times1.0, \times1.25, \times1.5$) of the standard change mode to the origin structure, producing a series of new structures. Then, we do SCF calculation to this series of structure and draw as figure3:
 \begin{figure}
  \centering
  \includegraphics[width=3.5in, height=5in]{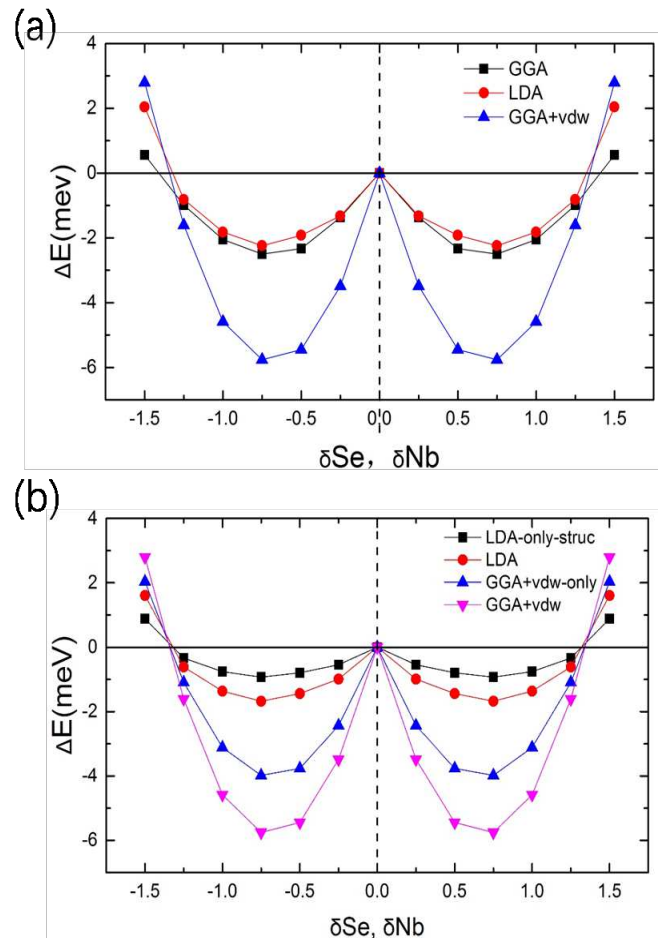}\\
  \caption{ (a) Shows the energy of a series new structure with different multiple oft the standard change mode by using different two exchange correlations function, GGA, LDA and mbj-GGA +vdw (including Van der Waals force). (b) shows the energy of two group structures with or without structure distortion (labeled by only-struc or not) calculated by two exchange correlations function, LDA and mbj-GGA +vdw (including Van der Waals force). X axis is the multiple index of atoms¡¯ movements; Y axis is the different value of their energy between the origin structure.}
  \label{fig3}
\end{figure}
Fig(3a) shows the result of three different calculation methods, GGA, LDA, and mbj-GGA considered Vander Waals forces. We can observe the energy have a strong dependence on exchange-correlation functional and the size of atoms ¡®displacements. And the largest different energy can reach 2meV lower than the origin structure by using mbj-GGA+vdw. For distinguish the effect from structure distortion and atom¡¯s displacement respectively, we produce two series of structures, one group only with structure distortion, the other one with two of them. And we do a SCF calculations to these two group and show out in figure 5.
According to Fig,3(b), all of these energy curves have the same trends, but energy will decrease more if you add structure distortion. Thus, structure distortion and atom¡¯s displacement can both lead to energy decreasing, which are the origin of CDW transition.
The analysis of energy successfully predicts the CDW instability and relationship between structure distortion and CDW transition. However, we have to make sure transition processes on the aspect of dynamics.

\begin{figure}[ht]
\centering
\subfigure[]{ \label{ff41}
 \includegraphics[width=1.5in]{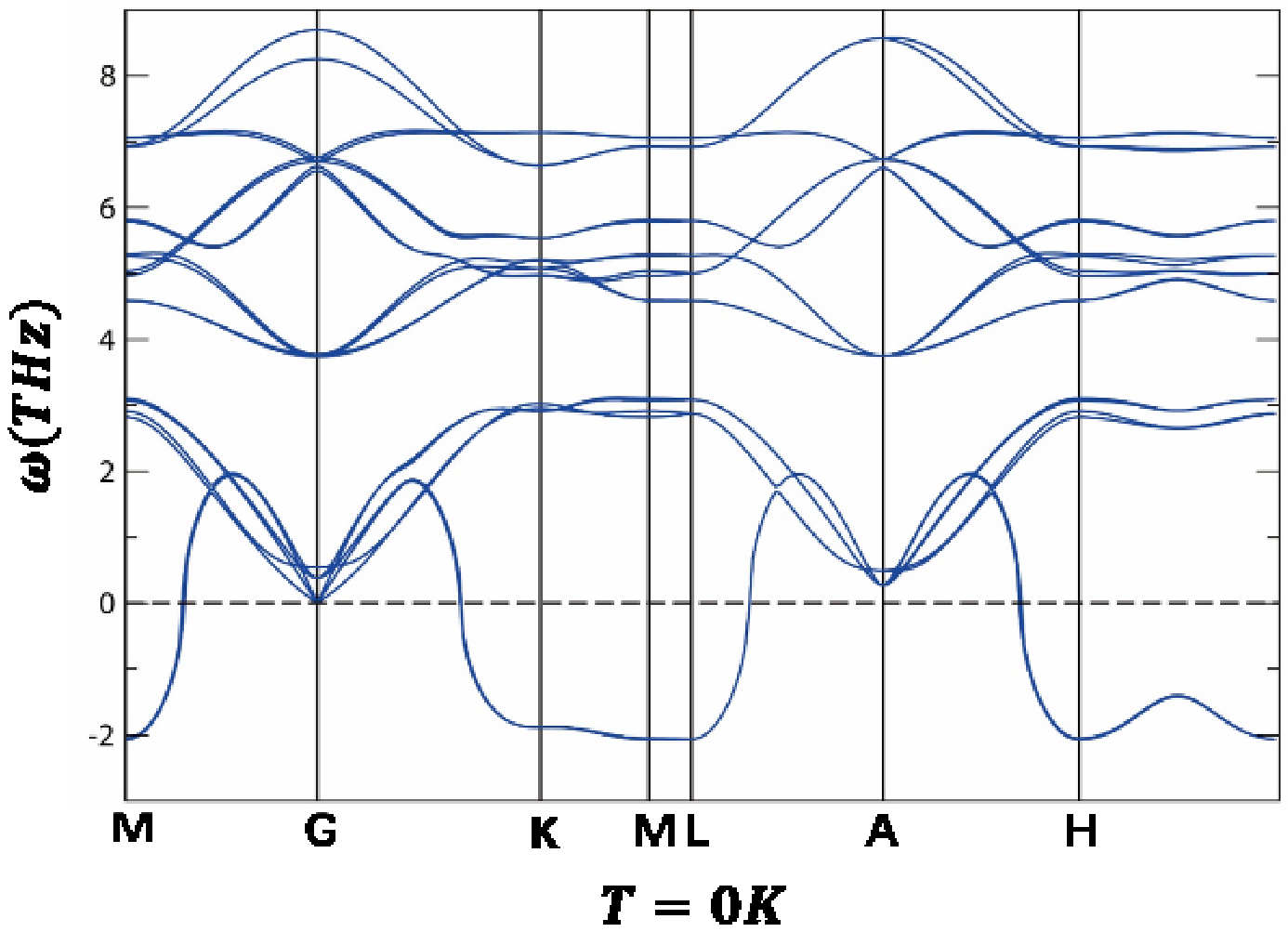}}

 \subfigure[]{ \label{ff42} 
 \includegraphics[width=1.5in]{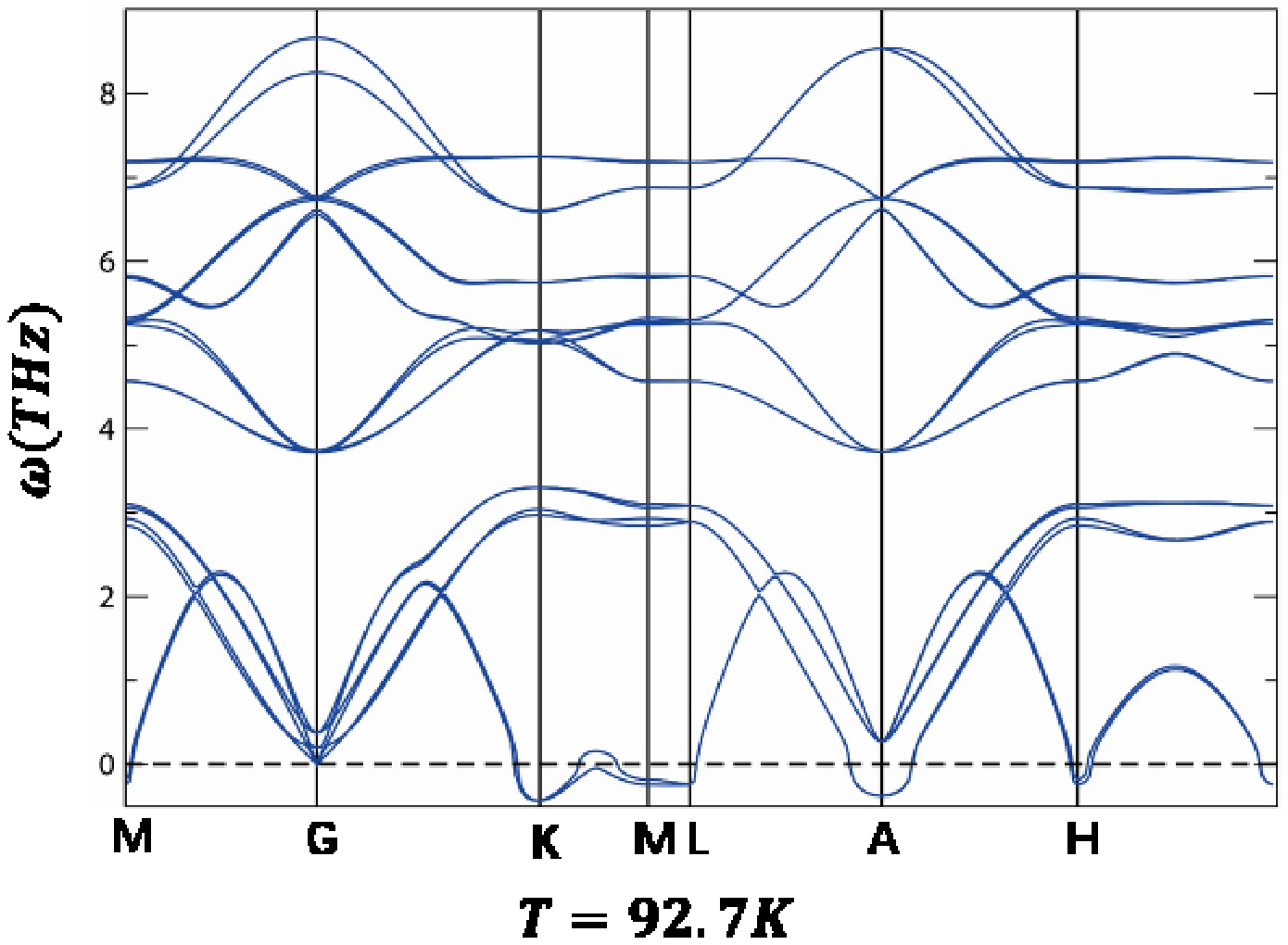}}
 \\
 \centering
 \subfigure[ ]{ \label{ff43}
 \includegraphics[width=1.5in]{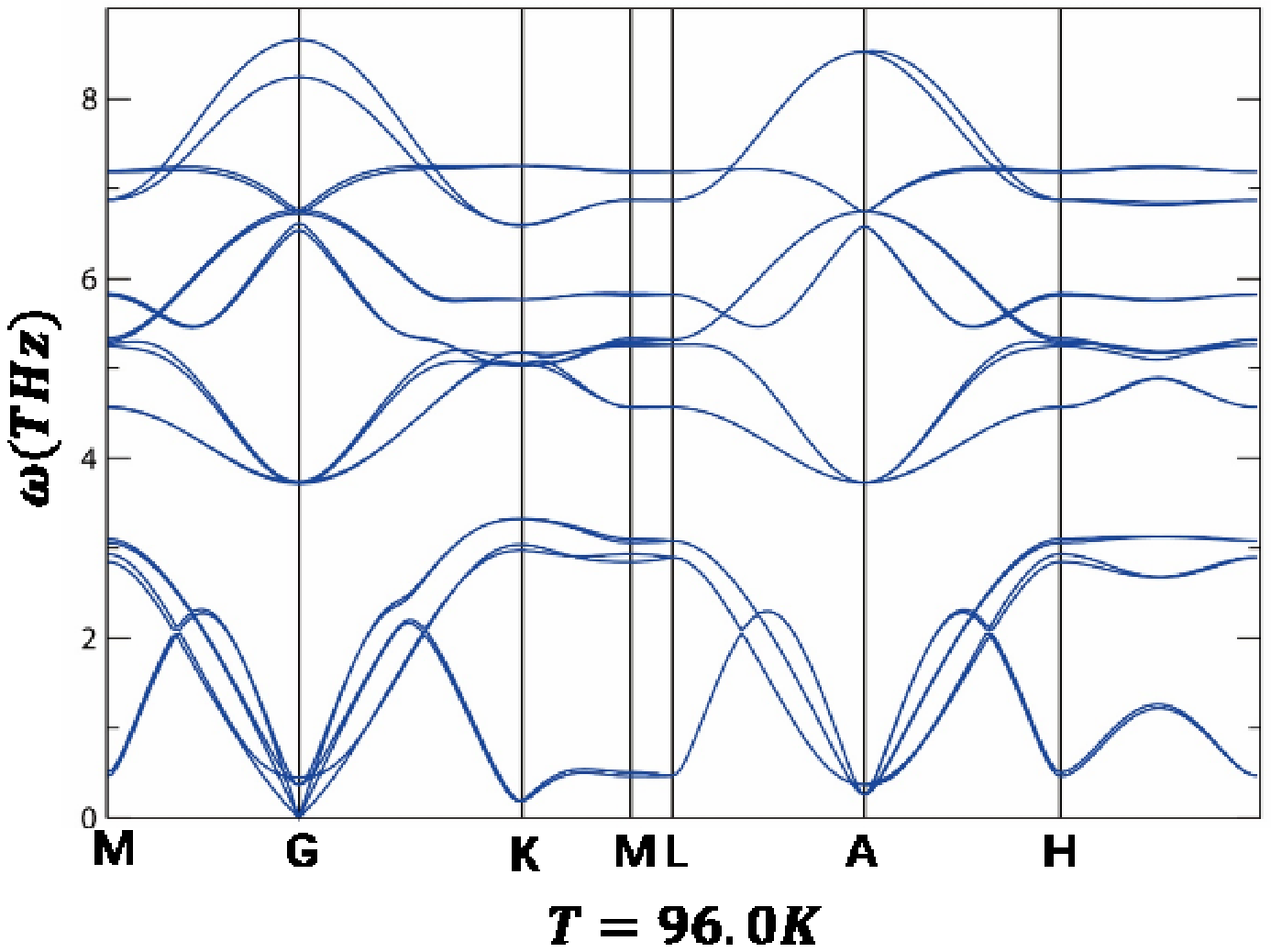}}

 \subfigure[]{ \label{ff44} 
 \includegraphics[width=1.5in]{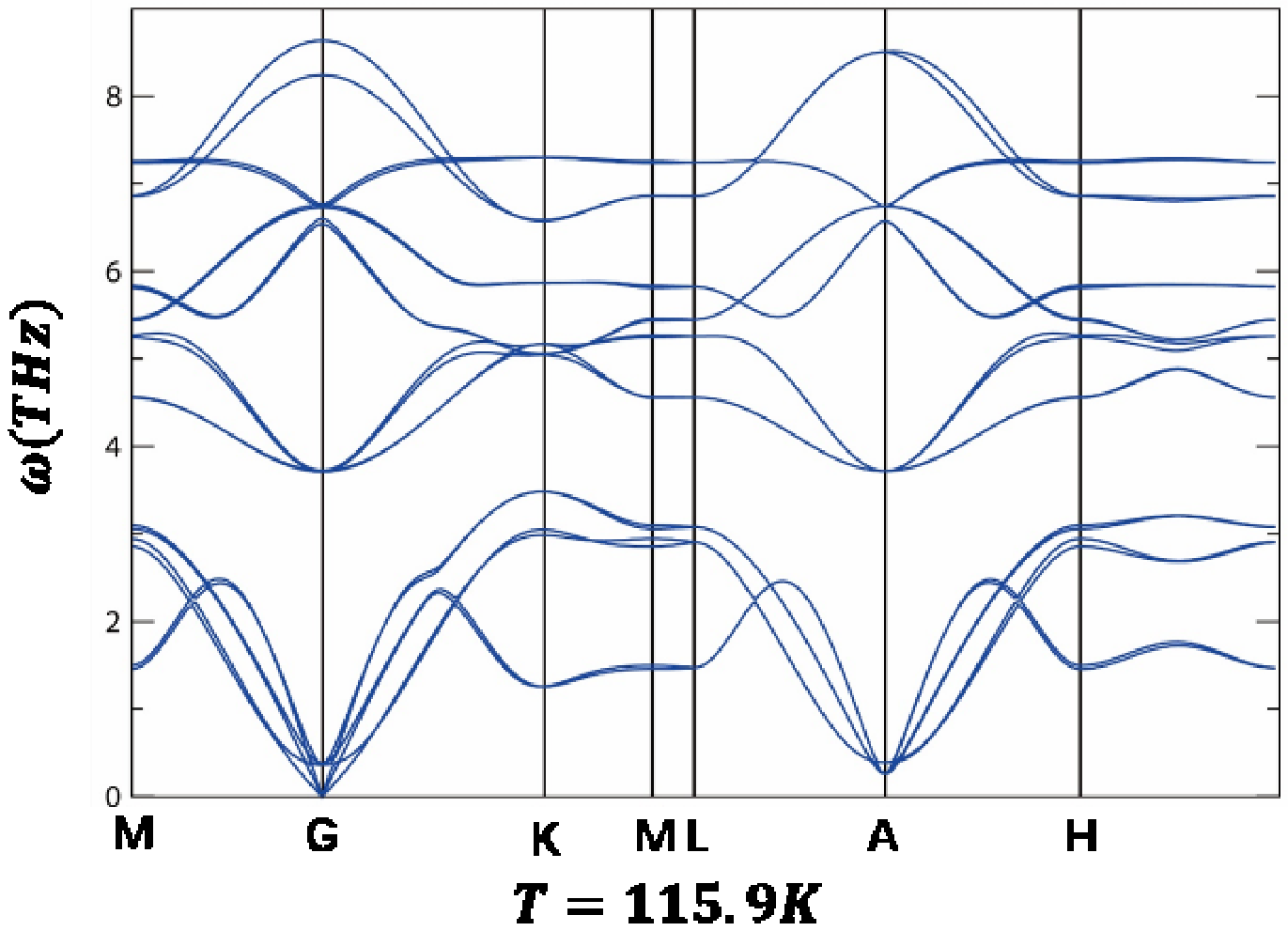}}

 \caption{Phonon dispersion spectrum of NbSe2 with space group $P6_3/mmc $ calculated by LDA functional and different Fermi smearing broadenings which correspond to the electron temperature. The electron temperature of each figure is written down under themselves. The x axis is position in K-space, and Y axis is the frequency of phonon($THz$), The symmetry points are labeled.  }
 \label{ff4}
\end{figure}
For investigate transition temperature of $NbSe_2$, the fermi smearing broadening of electrons is setting differently to simulate electronic dynamic of $NbSe_2$ under different temperature. According to Fig.~\ref{ff41}, even though $P6_3/mmc$ is the most stable space group for $NbSe_2$, it has imaginary frequency at K-M-L path when the corresponding temperature is under $96K$. And its imaginary frequency will decrease with temperature rising. It means that this structure is stable and unchanged at high temperature, and it may undergoes a transition at electron temperature $T_electron=96.0K$. Thus, we predict a transition in aspect of dynamics. Otherwise, we have to point out electrons temperature can¡¯t replace all of the effects caused by temperature, which is why our prediction, $T_electron=96.0K$, is lower than experimental result, $T_CDW=33.5K$.
If we want to prove our idea that the structure distortion and atom¡¯s displacement are the origin of CDW transition, we have to prove the appearing of charge density wave, beside energy and phonon. Thus, we calculate the charge properties of $NbSe_2$ and its change during CDW transition. The calculation was taken by VASP, with the LDA. Firstly, we calculate out the charge density of the origin order, called as $charge_1$, and CDW order, called as $charge_2$. (The detail of crystal structure is given before.) And we make subtraction $charge_s=charge_2-charge_1$, and show its distribution.

 \begin{figure}
  \centering
  \includegraphics[width=3.5in, height=2.0in]{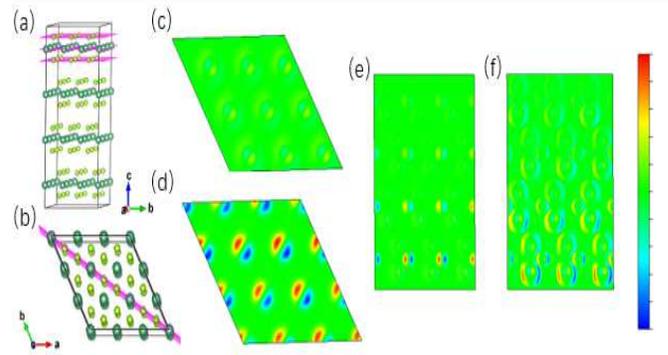}\\
  \caption{ Fig(a) and (b) show the positions two XY plane consisted with Nb and Se atoms and $110$ plane respectively, where fig(c)is for Se plane, and (d) for Nb plane. Both of (c) and (d) are 011 plane ($h=0, k=0, l=1$). Fig(c) shows the distribution of $charge_s$ on 2D plane with Nb atoms. And fig(d) show the distribution of $charge_s $on the plane with Se atoms. The red area means $charge_s>0$, where charge density increase during CDW transition. On the contrary, the blue area is where charge density decrease. Planes (e) and (f) are all 110 plane ($h=1, k=1, l=0$). Fig(e) and (f) show the distribution of the differential value of charge density ($charge_s$) and ELF value on 110 plane. The red area means the differential value $>0$, where charge density increase during CDW transition. On the contrary, the blue area is where charge density decrease.}
  \label{fig5}
\end{figure}

Figure.~\ref{fig5} tells us charge density¡¯s movements on the XY plane and along Z axis by the differential value of charge density and ELF(electronic localization function) . The movements on each layer are different, which is corresponding to the displacement of atoms basically.

 \begin{figure}
  \centering
  \includegraphics[width=3.5in, height=2.5in]{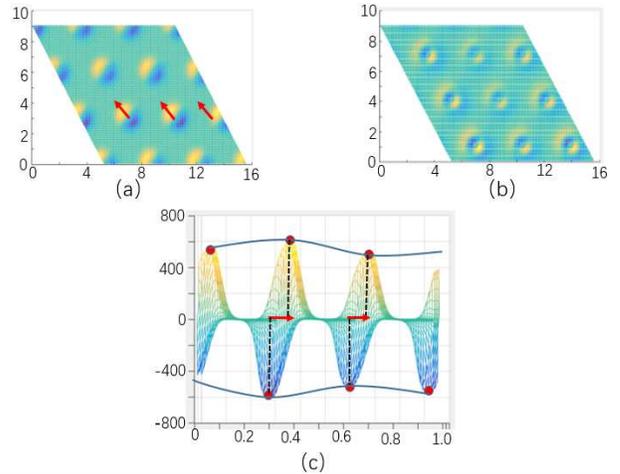}\\
  \caption{ (a) The density distribution of $charge_s$ on the XY plane with Nb atoms describing the change of charge density before and after CDW transition.  (b) The density distribution of $charge_s$ on the XY plane with Se plan. (c) the side view of fig(a) along X axis, where the amplitude max points are labeled with red points and connected by Sin function curve. The X axis is a crystal axis. And the Y axis is the value of $charge_s$. And $charge_s<0$ means this area loses charge, getting on the opposite. The red arrow shows the movements of charge density on X axis,}
  \label{fig6}
\end{figure}
The calculation in Fig.~\ref{fig5} and Fig.~\ref{fig6} are both taken in 3¡Á3¡Á2 supercell  with space group $P6_3/mmc$, meaning period is triple primary cell on X and Y axis. According to the side view in fig6(c), we can find that the charge density is modulated by Sin function based on the movement with atoms. So, the reason of periodic modulation should be CDW transition, which means the structure distortion and atoms¡¯displacment cause CDW transition. Expect charge density distribution, the results of ELF (electronic laicization function) calculation also prove that them are the origin of CDW transition.
 As for the period of CDW along z axis, there is no clear and certain conclusion of it. Thus, we observe its behavior along z axis showed in fig5(e) and fig5(f). The calculations are taken in $3\times3\times4$ supercell of $NbSe_2$ . According to the differential value of charge density before and after CDW transition on the 110 plane, we can find the amount of shifting charge is depending on the position on Z axis (deeper color means more charge¡¯s increasing in this area). Through analyzing, the charge density is also modulated by sin function, and its period is 4¡Áprimary cell. So the period of CDW in $NbSe_2$ should be $3\times3\times 4$.
We use a series of $NbSe_2' s$ structures which have large or small structure distortion and atoms¡¯displacement to prove that them are the origin of CDW transition, connected by electron-phonon coupling interactions. In fact, there are some other mechanisms, such as Fermi nesting and Saddle point. They both focus on the electronic density of state near fermi level, pointing out that the intensity of electron-phonon coupling increase on two parallel fermi surface connected by a certain vector $q_CDW$.
 \begin{figure}
  \centering
  \includegraphics[width=3.5in, height=2in]{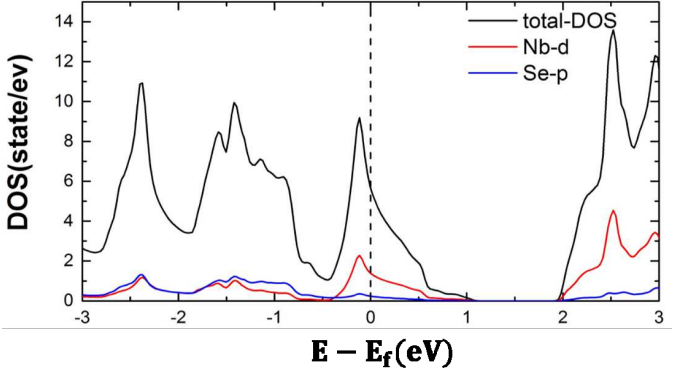}\\
  \caption{  The black line is total density of electronic state. The blue line is density of electronic state on Se $4p$ derived valence band, and red for Nb $4d$ band. X axis is the energy(eV) of electrons. Y axis is density of electronic state(state/eV). The fermi surface of $NbSe_2$ is almost driven by Nb $4d$ band.}
  \label{fig7}
\end{figure}

 \begin{figure}
  \centering
  \includegraphics[width=3.5in, height=1.4in]{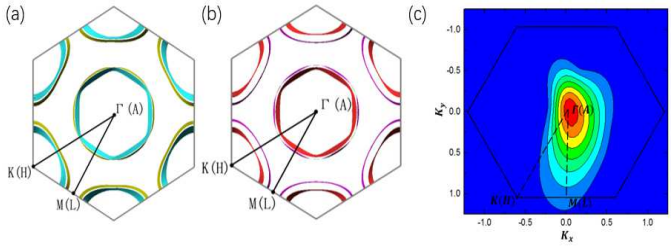}\\
  \caption{  (a) is the fermi surface of $NbSe_2$ without considering of the spin-orbit coupling(SOC) effect (b) shows the fermi surface without SOC. (c) is distribution of charge density in the plane $k_z=0$. X and Y axis are $k_x$ and $k_y$, respectively. }
  \label{fig8}
\end{figure}
 According to DOS(Fig.~\ref{fig7}) and band structure, we know $NbSe_2's$ fermi suface is basically consisted by Nb $4d$ and Se $4p$ band. Nb $4d$-derived conduction band displays a more important role. And the fermi cylindrical centered at$\Gamma A$ and $\Gamma M$ lead to fermi nesting where the density of electronic state is large. But for the nesting vector of the $\Gamma M$ cylinder, theoretic result is smaller than experiment result of ARPES. So, according to our result, there is not enough parallel  fermi surface and available electrons to form fermi nesting and lead to CDW transition. In addition, ref.\cite{Rice1975New} points out the distance of a pair of saddle points is incorrect, leading to a wrong $2\times2$ superstructure. In a word, these two mechanism are not the essential origins of CDW transition. To prove our ideas, how structure distortion and atom¡¯s displacement affect $NbSe_2' s$ properties in K space is supposed to be shown out.
First of all, we have to recognize that the effect taken by CDW transition is tiny no matter what kind of perspective you view from, because $k_BT_CDW=1.38\times10^{-23} (J/K)\times33.5K\approx2.89meV$. Thus the effect caused by structure distortion and atom¡¯s displacement is also tiny. But the distribution of electronic density in Kspace will happen to change.
 \begin{figure}
  \centering
  \includegraphics[width=3.5in, height=1.7in]{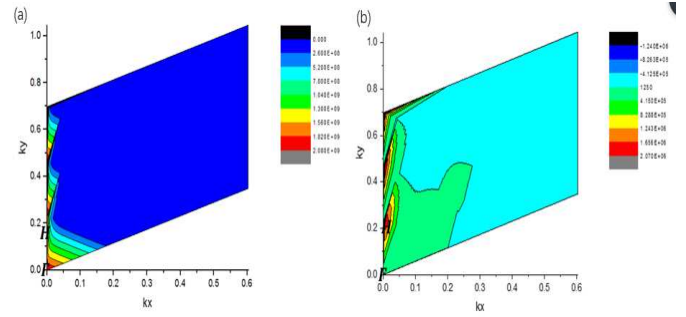}\\
  \caption{ (a) describes the charge density on the plane, $k_z=0$. And the right figure show the differential value of charge density on the plane $k_z=0$ of two structure with distortion and atom¡¯s displacement. }
  \label{fig9}
\end{figure}
According to fig.~\ref{fig9}, we can find that the charge density concentrate on $\Gamma(A)¡¢H(K)$ and $M(L)$ in supercell of $NbSe_2$. Beside this, charge density at $H$ and $K$ increase a lot after the structure distortion and atom¡¯s displacement happening, which leads to more available charge on fermi nesting cylindrical surfaces.  Thus, the CDW vector should depend on fermi nesting on $\Gamma M$,  $\Gamma K$ and self-nesting centered at $\Gamma A$. And its lower limit is $0.603 A^{-1}$, which is closer with the experimental value $|q_CDW=0.688A^{-1} |$ than some other papers¡¯
Combing with the content before, we may draw a conclusion that the structure distortion and atom¡¯s displacement might be an essential factor of CDW transition, because we can observed the charge density waves happen in $NbSe_2$ after structure distortion and atoms¡¯ displacement.
\section{The Relationship between SC and CDW in $NbSe_2$}
  In addition, the attractive points of $NbSe_2$ is not only CDW transition and its origin, but also the relationship between superconducting order and CDW order. So we also investigate it and explain the related behaviors from the aspect of structure distortion and atoms¡¯ displacement in supercell. As a lot of experiment reporting\cite{langer2014giant,Berthier1976Evidence}, the CDW order and SC order can coexist and complicate with each other in $NbSe_2$ at temperature range of $0-10K$. If the external pressures is rising by applying an hydrostatic pressure experimentally , $T_CDW $will decrease and the CDW instability will also disappear above the critical pressure $P_CDW=4.6GPa$ (when $T=3.5K$). It means the relationship between CDW order and SC order is competition. Through simple analysis, the formation of CDW and SC order all need the electrons for coupling near fermi level with energy gap, which is finite.
In $2H-NbSe_2$, the density of electronic state in path $\Gamma-K-H$ on fermi surface is high. It can be used for different kinds of electron pairing leading to SC or CDW transition respectively. However, fermi surface centered at ¦£ point lacks of energy gap for CDW transition actually. Thus, the strength of CDW and SC order is also decided by the characters of fermi nesting and electricity near K point, which requires that the distance of parallel fermi cylinders is matched to $q_CDW$. In fact, the requirement of CDW is harder to fit than SC¡¯s. So the transition temperature of CDW will decrease with pressure increasing, but SC¡¯s is still a constant.
 \begin{figure}
  \centering
  \includegraphics[width=3.5in, height=1.8in]{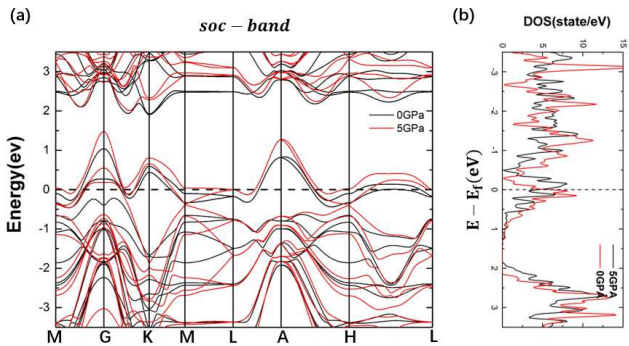}\\
  \caption{  (a) is the schematic figure of $NbSe_2's$ band structure at pressure of $0GPa$ and $5GPa$, where the spin-orbit coupling effect is taken into considering and calculation. (b) shows the density of electronic state (DOS). The Y axis is DOS($state/ev$), and the X axis is the value of $E-E_f (eV)$, range of $-3.5eV~3.5eV$ corresponding to the energy range of band. }
  \label{fig10}
\end{figure}
According to Fig.~\ref{fig10}, the energy gap decreases at $\Gamma$ point, and at K point slightly following pressure¡¯s increase. Meanwhile, the density of electronic state decrease at fermi level meaning no too much appropriate electrons for pairing. Both of these two points result in that the rest electrons can only be used as cooper pair but not CDW order, which is identical to the experimental result.
\section{Conclusion}
In this paper, we investigated structures and charge density waves, including its origin of CDW, by using first-principles calculations. At First, we performed structure searches of $NbSe_2$ to explore the possible structures of the CDW phase in $NbSe_2$. To prove the stability of the resulted structure, we do extensive studies of energy calculations, phonon spectra and other calculations. As for the origin of CDW, We believe that it¡¯s the structure distortion and the displacement of atoms. And we analyze the electronic properties (both in real space and K space) and band structures during CDW transitions. In the end, we perform the result of band structure and DOS under different temperature, and explain the reason why CDW and superconducting orders compete with each other at low temperature.
\nocite{*}

\bibliography{apssamp}

\begin{thebibliography}{18}%
\makeatletter
\providecommand \@ifxundefined [1]{%
 \@ifx{#1\undefined}
}%
\providecommand \@ifnum [1]{%
 \ifnum #1\expandafter \@firstoftwo
 \else \expandafter \@secondoftwo
 \fi
}%
\providecommand \@ifx [1]{%
 \ifx #1\expandafter \@firstoftwo
 \else \expandafter \@secondoftwo
 \fi
}%
\providecommand \natexlab [1]{#1}%
\providecommand \enquote  [1]{``#1''}%
\providecommand \bibnamefont  [1]{#1}%
\providecommand \bibfnamefont [1]{#1}%
\providecommand \citenamefont [1]{#1}%
\providecommand \href@noop [0]{\@secondoftwo}%
\providecommand \href [0]{\begingroup \@sanitize@url \@href}%
\providecommand \@href[1]{\@@startlink{#1}\@@href}%
\providecommand \@@href[1]{\endgroup#1\@@endlink}%
\providecommand \@sanitize@url [0]{\catcode `\\12\catcode `\$12\catcode
  `\&12\catcode `\#12\catcode `\^12\catcode `\_12\catcode `\%12\relax}%
\providecommand \@@startlink[1]{}%
\providecommand \@@endlink[0]{}%
\providecommand \url  [0]{\begingroup\@sanitize@url \@url }%
\providecommand \@url [1]{\endgroup\@href {#1}{\urlprefix }}%
\providecommand \urlprefix  [0]{URL }%
\providecommand \Eprint [0]{\href }%
\providecommand \doibase [0]{http://dx.doi.org/}%
\providecommand \selectlanguage [0]{\@gobble}%
\providecommand \bibinfo  [0]{\@secondoftwo}%
\providecommand \bibfield  [0]{\@secondoftwo}%
\providecommand \translation [1]{[#1]}%
\providecommand \BibitemOpen [0]{}%
\providecommand \bibitemStop [0]{}%
\providecommand \bibitemNoStop [0]{.\EOS\space}%
\providecommand \EOS [0]{\spacefactor3000\relax}%
\providecommand \BibitemShut  [1]{\csname bibitem#1\endcsname}%
\let\auto@bib@innerbib\@empty
\bibitem [{\citenamefont {Withers}\ and\ \citenamefont
  {Wilson}(1986)}]{Withers1986REVIEW}%
  \BibitemOpen
  \bibfield  {author} {\bibinfo {author} {\bibfnamefont {R.~L.}\ \bibnamefont
  {Withers}}\ and\ \bibinfo {author} {\bibfnamefont {J.~A.}\ \bibnamefont
  {Wilson}},\ }\href@noop {} {\bibfield  {journal} {\bibinfo  {journal}
  {Journal of Physics C Solid State Physics}\ }\textbf {\bibinfo {volume} {19}}
  (\bibinfo {year} {1986})}\BibitemShut {NoStop}%
\bibitem [{\citenamefont {Berggren}\ and\ \citenamefont
  {Pepper}(2010)}]{Berggren2010Electrons}%
  \BibitemOpen
  \bibfield  {author} {\bibinfo {author} {\bibfnamefont {K.~F.}\ \bibnamefont
  {Berggren}}\ and\ \bibinfo {author} {\bibfnamefont {M.}~\bibnamefont
  {Pepper}},\ }\href@noop {} {\bibfield  {journal} {\bibinfo  {journal} {Philos
  Trans A Math Phys Eng Sci}\ }\textbf {\bibinfo {volume} {368}},\ \bibinfo
  {pages} {1141} (\bibinfo {year} {2010})}\BibitemShut {NoStop}%
\bibitem [{\citenamefont {Bardeen}(1990)}]{Bardeen1990Superconductivity}%
  \BibitemOpen
  \bibfield  {author} {\bibinfo {author} {\bibfnamefont {J.}~\bibnamefont
  {Bardeen}},\ }\href@noop {} {\bibfield  {journal} {\bibinfo  {journal}
  {Physics Today}\ }\textbf {\bibinfo {volume} {43}},\ \bibinfo {pages} {25}
  (\bibinfo {year} {1990})}\BibitemShut {NoStop}%
\bibitem [{\citenamefont {Wilson}\ \emph {et~al.}(2001)\citenamefont {Wilson},
  \citenamefont {Salvo},\ and\ \citenamefont {Mahajan}}]{J2001Charge}%
  \BibitemOpen
  \bibfield  {author} {\bibinfo {author} {\bibfnamefont {J.}~\bibnamefont
  {Wilson}}, \bibinfo {author} {\bibfnamefont {F.~D.}\ \bibnamefont {Salvo}}, \
  and\ \bibinfo {author} {\bibfnamefont {S.}~\bibnamefont {Mahajan}},\
  }\href@noop {} {\bibfield  {journal} {\bibinfo  {journal} {Advances in
  Physics}\ }\textbf {\bibinfo {volume} {50}},\ \bibinfo {pages} {1171}
  (\bibinfo {year} {2001})}\BibitemShut {NoStop}%
\bibitem [{\citenamefont {Langer}\ \emph {et~al.}(2014)\citenamefont {Langer},
  \citenamefont {Kisiel}, \citenamefont {Pawlak}, \citenamefont {Pellegrini},
  \citenamefont {Santoro}, \citenamefont {Buzio}, \citenamefont {Gerbi},
  \citenamefont {Balakrishnan}, \citenamefont {Baratoff},\ and\ \citenamefont
  {Tosatti}}]{Langer2014Giant}%
  \BibitemOpen
  \bibfield  {author} {\bibinfo {author} {\bibfnamefont {M.}~\bibnamefont
  {Langer}}, \bibinfo {author} {\bibfnamefont {M.}~\bibnamefont {Kisiel}},
  \bibinfo {author} {\bibfnamefont {R.}~\bibnamefont {Pawlak}}, \bibinfo
  {author} {\bibfnamefont {F.}~\bibnamefont {Pellegrini}}, \bibinfo {author}
  {\bibfnamefont {G.~E.}\ \bibnamefont {Santoro}}, \bibinfo {author}
  {\bibfnamefont {R.}~\bibnamefont {Buzio}}, \bibinfo {author} {\bibfnamefont
  {A.}~\bibnamefont {Gerbi}}, \bibinfo {author} {\bibfnamefont
  {G.}~\bibnamefont {Balakrishnan}}, \bibinfo {author} {\bibfnamefont
  {A.}~\bibnamefont {Baratoff}}, \ and\ \bibinfo {author} {\bibfnamefont
  {E.}~\bibnamefont {Tosatti}},\ }\href@noop {} {\bibfield  {journal} {\bibinfo
   {journal} {Nature Materials}\ }\textbf {\bibinfo {volume} {13}},\ \bibinfo
  {pages} {173} (\bibinfo {year} {2014})}\BibitemShut {NoStop}%
\bibitem [{\citenamefont {Johannes}\ and\ \citenamefont
  {Mazin}(2007)}]{M2007Fermi}%
  \BibitemOpen
  \bibfield  {author} {\bibinfo {author} {\bibfnamefont {M.}~\bibnamefont
  {Johannes}}\ and\ \bibinfo {author} {\bibfnamefont {I.~I.}\ \bibnamefont
  {Mazin}},\ }\href@noop {} {\bibfield  {journal} {\bibinfo  {journal}
  {Physical Review B Condensed Matter}\ }\textbf {\bibinfo {volume} {77}},\
  (\bibinfo {year} {2007})}\BibitemShut {NoStop}%
\bibitem [{\citenamefont {Gr¨¹ner}(1994)}]{Gr1994Density}%
  \BibitemOpen
  \bibfield  {author} {\bibinfo {author} {\bibfnamefont {G.}~\bibnamefont
  {Gr¨¹ner}},\ }\href@noop {} {\bibfield  {journal} {\bibinfo  {journal}
  {Frontiers in Physics}\ } (\bibinfo {year} {1994})}\BibitemShut {NoStop}%
\bibitem [{\citenamefont {Straub}\ \emph {et~al.}(1999)\citenamefont {Straub},
  \citenamefont {Finteis}, \citenamefont {Claessen}, \citenamefont {Steiner},
  \citenamefont {H¨¹fner}, \citenamefont {Blaha}, \citenamefont {Oglesby},\
  and\ \citenamefont {Bucher}}]{Straub1999Charge}%
  \BibitemOpen
  \bibfield  {author} {\bibinfo {author} {\bibfnamefont {T.}~\bibnamefont
  {Straub}}, \bibinfo {author} {\bibfnamefont {T.}~\bibnamefont {Finteis}},
  \bibinfo {author} {\bibfnamefont {R.}~\bibnamefont {Claessen}}, \bibinfo
  {author} {\bibfnamefont {P.}~\bibnamefont {Steiner}}, \bibinfo {author}
  {\bibfnamefont {S.}~\bibnamefont {H¨¹fner}}, \bibinfo {author} {\bibfnamefont
  {P.}~\bibnamefont {Blaha}}, \bibinfo {author} {\bibfnamefont {C.~S.}\
  \bibnamefont {Oglesby}}, \ and\ \bibinfo {author} {\bibfnamefont
  {E.}~\bibnamefont {Bucher}},\ }\href@noop {} {\bibfield  {journal} {\bibinfo
  {journal} {Phys.rev.lett}\ }\textbf {\bibinfo {volume} {82}},\ \bibinfo
  {pages} {4504} (\bibinfo {year} {1999})}\BibitemShut {NoStop}%
\bibitem [{\citenamefont {Rice}\ and\ \citenamefont
  {Scott}(1975)}]{Rice1975New}%
  \BibitemOpen
  \bibfield  {author} {\bibinfo {author} {\bibfnamefont {T.~M.}\ \bibnamefont
  {Rice}}\ and\ \bibinfo {author} {\bibfnamefont {G.~K.}\ \bibnamefont
  {Scott}},\ }\href@noop {} {\bibfield  {journal} {\bibinfo  {journal}
  {Physical Review Letters}\ }\textbf {\bibinfo {volume} {35}},\ \bibinfo
  {pages} {120} (\bibinfo {year} {1975})}\BibitemShut {NoStop}%
\bibitem [{\citenamefont {Valla}\ \emph {et~al.}(2000)\citenamefont {Valla},
  \citenamefont {Fedorov}, \citenamefont {Johnson}, \citenamefont {Xue},
  \citenamefont {Smith},\ and\ \citenamefont {Disalvo}}]{Valla2000Charge}%
  \BibitemOpen
  \bibfield  {author} {\bibinfo {author} {\bibfnamefont {T.}~\bibnamefont
  {Valla}}, \bibinfo {author} {\bibfnamefont {A.~V.}\ \bibnamefont {Fedorov}},
  \bibinfo {author} {\bibfnamefont {P.~D.}\ \bibnamefont {Johnson}}, \bibinfo
  {author} {\bibfnamefont {J.}~\bibnamefont {Xue}}, \bibinfo {author}
  {\bibfnamefont {K.~E.}\ \bibnamefont {Smith}}, \ and\ \bibinfo {author}
  {\bibfnamefont {F.~J.}\ \bibnamefont {Disalvo}},\ }\href@noop {} {\bibfield
  {journal} {\bibinfo  {journal} {Physical Review Letters}\ }\textbf {\bibinfo
  {volume} {85}},\ \bibinfo {pages} {4759} (\bibinfo {year}
  {2000})}\BibitemShut {NoStop}%
\bibitem [{\citenamefont {Kiss}\ \emph {et~al.}(2007)\citenamefont {Kiss},
  \citenamefont {Yokoya}, \citenamefont {Chainani}, \citenamefont {Shin},
  \citenamefont {Hanaguri}, \citenamefont {Nohara},\ and\ \citenamefont
  {Takagi}}]{Kiss2007Charge}%
  \BibitemOpen
  \bibfield  {author} {\bibinfo {author} {\bibfnamefont {T.}~\bibnamefont
  {Kiss}}, \bibinfo {author} {\bibfnamefont {T.}~\bibnamefont {Yokoya}},
  \bibinfo {author} {\bibfnamefont {A.}~\bibnamefont {Chainani}}, \bibinfo
  {author} {\bibfnamefont {S.}~\bibnamefont {Shin}}, \bibinfo {author}
  {\bibfnamefont {T.}~\bibnamefont {Hanaguri}}, \bibinfo {author}
  {\bibfnamefont {M.}~\bibnamefont {Nohara}}, \ and\ \bibinfo {author}
  {\bibfnamefont {H.}~\bibnamefont {Takagi}},\ }\href@noop {} {\bibfield
  {journal} {\bibinfo  {journal} {Nature Physics}\ }\textbf {\bibinfo {volume}
  {3}},\ \bibinfo {pages} {720} (\bibinfo {year} {2007})}\BibitemShut {NoStop}%
\bibitem [{\citenamefont {Weber}\ \emph {et~al.}(2011)\citenamefont {Weber},
  \citenamefont {Rosenkranz}, \citenamefont {Castellan}, \citenamefont
  {Osborn}, \citenamefont {Hott}, \citenamefont {Heid}, \citenamefont {Bohnen},
  \citenamefont {Egami}, \citenamefont {Said},\ and\ \citenamefont
  {Reznik}}]{Weber2011Extended}%
  \BibitemOpen
  \bibfield  {author} {\bibinfo {author} {\bibfnamefont {F.}~\bibnamefont
  {Weber}}, \bibinfo {author} {\bibfnamefont {S.}~\bibnamefont {Rosenkranz}},
  \bibinfo {author} {\bibfnamefont {J.~P.}\ \bibnamefont {Castellan}}, \bibinfo
  {author} {\bibfnamefont {R.}~\bibnamefont {Osborn}}, \bibinfo {author}
  {\bibfnamefont {R.}~\bibnamefont {Hott}}, \bibinfo {author} {\bibfnamefont
  {R.}~\bibnamefont {Heid}}, \bibinfo {author} {\bibfnamefont {K.~P.}\
  \bibnamefont {Bohnen}}, \bibinfo {author} {\bibfnamefont {T.}~\bibnamefont
  {Egami}}, \bibinfo {author} {\bibfnamefont {A.~H.}\ \bibnamefont {Said}}, \
  and\ \bibinfo {author} {\bibfnamefont {D.}~\bibnamefont {Reznik}},\
  }\href@noop {} {\bibfield  {journal} {\bibinfo  {journal} {Physical Review
  Letters}\ }\textbf {\bibinfo {volume} {107}},\ \bibinfo {pages} {737}
  (\bibinfo {year} {2011})}\BibitemShut {NoStop}%
\bibitem [{\citenamefont {Arguello}\ \emph {et~al.}(2015)\citenamefont
  {Arguello}, \citenamefont {Rosenthal}, \citenamefont {Andrade}, \citenamefont
  {Jin}, \citenamefont {Yeh}, \citenamefont {Zaki}, \citenamefont {Jia},
  \citenamefont {Cava}, \citenamefont {Fernandes},\ and\ \citenamefont
  {Millis}}]{C2015Quasiparticle}%
  \BibitemOpen
  \bibfield  {author} {\bibinfo {author} {\bibfnamefont {C.}~\bibnamefont
  {Arguello}}, \bibinfo {author} {\bibfnamefont {E.}~\bibnamefont {Rosenthal}},
  \bibinfo {author} {\bibfnamefont {E.}~\bibnamefont {Andrade}}, \bibinfo
  {author} {\bibfnamefont {W.}~\bibnamefont {Jin}}, \bibinfo {author}
  {\bibfnamefont {P.}~\bibnamefont {Yeh}}, \bibinfo {author} {\bibfnamefont
  {N.}~\bibnamefont {Zaki}}, \bibinfo {author} {\bibfnamefont {S.}~\bibnamefont
  {Jia}}, \bibinfo {author} {\bibfnamefont {R.}~\bibnamefont {Cava}}, \bibinfo
  {author} {\bibfnamefont {R.}~\bibnamefont {Fernandes}}, \ and\ \bibinfo
  {author} {\bibfnamefont {A.}~\bibnamefont {Millis}},\ }\href@noop {}
  {\bibfield  {journal} {\bibinfo  {journal} {Physical Review Letters}\
  }\textbf {\bibinfo {volume} {114}},\ \bibinfo {pages} {037001} (\bibinfo
  {year} {2015})}\BibitemShut {NoStop}%
\bibitem [{\citenamefont {Soto}\ \emph {et~al.}(2006)\citenamefont {Soto},
  \citenamefont {Berger}, \citenamefont {Cabo}, \citenamefont {Carballeira},
  \citenamefont {Mosqueira}, \citenamefont {Pavuna}, \citenamefont {Toimil},\
  and\ \citenamefont {Vidal}}]{Soto2006Electric}%
  \BibitemOpen
  \bibfield  {author} {\bibinfo {author} {\bibfnamefont {F.}~\bibnamefont
  {Soto}}, \bibinfo {author} {\bibfnamefont {H.}~\bibnamefont {Berger}},
  \bibinfo {author} {\bibfnamefont {L.}~\bibnamefont {Cabo}}, \bibinfo {author}
  {\bibfnamefont {C.}~\bibnamefont {Carballeira}}, \bibinfo {author}
  {\bibfnamefont {J.}~\bibnamefont {Mosqueira}}, \bibinfo {author}
  {\bibfnamefont {D.}~\bibnamefont {Pavuna}}, \bibinfo {author} {\bibfnamefont
  {P.}~\bibnamefont {Toimil}}, \ and\ \bibinfo {author} {\bibfnamefont
  {F.}~\bibnamefont {Vidal}},\ }\href@noop {} {\bibfield  {journal} {\bibinfo
  {journal} {Physica C Superconductivity}\ }\textbf {\bibinfo {volume} {460}},\
  \bibinfo {pages} {789} (\bibinfo {year} {2006})}\BibitemShut {NoStop}%
\bibitem [{\citenamefont {Leroux}\ \emph {et~al.}(2015)\citenamefont {Leroux},
  \citenamefont {Errea}, \citenamefont {Tacon}, \citenamefont {Souliou},
  \citenamefont {Garbarino}, \citenamefont {Cario}, \citenamefont {Bosak},
  \citenamefont {Mauri}, \citenamefont {Calandra},\ and\ \citenamefont
  {Rodi¨¨re}}]{Leroux2015Strong}%
  \BibitemOpen
  \bibfield  {author} {\bibinfo {author} {\bibfnamefont {M.}~\bibnamefont
  {Leroux}}, \bibinfo {author} {\bibfnamefont {I.}~\bibnamefont {Errea}},
  \bibinfo {author} {\bibfnamefont {M.~L.}\ \bibnamefont {Tacon}}, \bibinfo
  {author} {\bibfnamefont {S.~M.}\ \bibnamefont {Souliou}}, \bibinfo {author}
  {\bibfnamefont {G.}~\bibnamefont {Garbarino}}, \bibinfo {author}
  {\bibfnamefont {L.}~\bibnamefont {Cario}}, \bibinfo {author} {\bibfnamefont
  {A.}~\bibnamefont {Bosak}}, \bibinfo {author} {\bibfnamefont
  {F.}~\bibnamefont {Mauri}}, \bibinfo {author} {\bibfnamefont
  {M.}~\bibnamefont {Calandra}}, \ and\ \bibinfo {author} {\bibfnamefont
  {P.}~\bibnamefont {Rodi¨¨re}},\ }\href@noop {} {\bibfield  {journal}
  {\bibinfo  {journal} {Phys.rev.b}\ }\textbf {\bibinfo {volume} {92}},\
  \bibinfo {pages} {140303} (\bibinfo {year} {2015})}\BibitemShut {NoStop}%
\bibitem [{\citenamefont {Burke}\ \emph {et~al.}(1998)\citenamefont {Burke},
  \citenamefont {Perdew},\ and\ \citenamefont {Wang}}]{Burke1998Derivation}%
  \BibitemOpen
  \bibfield  {author} {\bibinfo {author} {\bibfnamefont {K.}~\bibnamefont
  {Burke}}, \bibinfo {author} {\bibfnamefont {J.~P.}\ \bibnamefont {Perdew}}, \
  and\ \bibinfo {author} {\bibfnamefont {Y.}~\bibnamefont {Wang}},\ }\href@noop
  {} {\emph {\bibinfo {title} {Derivation of a Generalized Gradient
  Approximation: The PW91 Density Functional}}}\ (\bibinfo  {publisher}
  {Springer US},\ \bibinfo {year} {1998})\ pp.\ \bibinfo {pages}
  {81--111}\BibitemShut {NoStop}%
\bibitem [{\citenamefont {Perdew}\ \emph {et~al.}(1996)\citenamefont {Perdew},
  \citenamefont {Burke},\ and\ \citenamefont
  {Ernzerhof}}]{perdew1996generalized}%
  \BibitemOpen
  \bibfield  {author} {\bibinfo {author} {\bibfnamefont {J.~P.}\ \bibnamefont
  {Perdew}}, \bibinfo {author} {\bibfnamefont {K.}~\bibnamefont {Burke}}, \
  and\ \bibinfo {author} {\bibfnamefont {M.}~\bibnamefont {Ernzerhof}},\
  }\href@noop {} {\bibfield  {journal} {\bibinfo  {journal} {Physical review
  letters}\ }\textbf {\bibinfo {volume} {77}},\ \bibinfo {pages} {3865}
  (\bibinfo {year} {1996})}\BibitemShut {NoStop}%
\bibitem [{\citenamefont {Berthier}\ \emph {et~al.}(1976)\citenamefont
  {Berthier}, \citenamefont {Molini¨¦},\ and\ \citenamefont
  {J¨¦rome}}]{Berthier1976Evidence}%
  \BibitemOpen
  \bibfield  {author} {\bibinfo {author} {\bibfnamefont {C.}~\bibnamefont
  {Berthier}}, \bibinfo {author} {\bibfnamefont {P.}~\bibnamefont {Molini¨¦}},
  \ and\ \bibinfo {author} {\bibfnamefont {D.}~\bibnamefont {J¨¦rome}},\
  }\href@noop {} {\bibfield  {journal} {\bibinfo  {journal} {Solid State
  Communications}\ }\textbf {\bibinfo {volume} {18}},\ \bibinfo {pages} {1393}
  (\bibinfo {year} {1976})}\BibitemShut {NoStop}%
\end{thebibliography}%

\end{document}